\documentclass[aps,prl,twocolumn]{revtex4}

 \def\be{\begin{eqnarray}}
\def\ee{\end{eqnarray}}

\usepackage{graphicx}

\begin{document}

\title{Connecting Lattice QCD with Chiral Perturbation Theory at 
Strong Coupling}
\author{Shailesh Chandrasekharan$^a$ and Costas G. Strouthos$^{a,b}$}
\affiliation{
$^a$Department of Physics, Box 90305, Duke University,
Durham, North Carolina 27708. \\
$^b$Division of Science and Engineering, Frederick Institute of Technology,
Nicosia 1303, Cyprus.}

\preprint{DUKE-TH-03-259}

\begin{abstract}
We study the difficulties associated with detecting chiral 
singularities predicted by chiral perturbation theory (ChPT) in lattice 
QCD. We focus on the physics of the remnant $O(2)$ chiral 
symmetry of staggered fermions in the strong coupling limit using the
recently discovered directed path algorithm. Since it is easier to look 
for power-like singularities as compared to logarithmic ones, our 
calculations are performed at a fixed finite temperature in the
chirally broken phase.  We show that the behavior 
of the chiral condensate, the pion mass and the pion decay constant, for 
small masses, are all consistent with the predictions of ChPT.
 However, the values of the quark masses that we need to demonstrate 
this are much smaller than those being used in dynamical QCD simulations. 
We also need to use higher order terms in the chiral expansion to fit
our data.

\end{abstract}

\maketitle

\section{INTRODUCTION}

With the advent of fast computers it is now possible to calculate
many hadronic quantities from first principles using lattice QCD.
However, today these calculations contain systematic errors due to
finite lattice spacings, finite volumes and quenching. Although in 
principle all these errors can be controlled, the only clean way
to reduce quenching errors is to perform unquenched calculations.
Due to algorithmic difficulties today most unquenched calculations 
are performed using large quark masses and the results are then 
extrapolated to realistic values using (quenched and partially quenched) 
ChPT. Unfortunately, recent attempts to connect lattice QCD with 
the usual one-loop ChPT predictions have failed to give 
clear answers \cite{JLQCD,Dur02,Mon03,Ber02}. It is now believed that 
lattice artifacts should be taken into consideration in ChPT
\cite{Lee99,Aub03,Rup02,Aok03}. It has been suggested in \cite{Aub03.a,Far03} 
the lattice data is described better by the resulting more elaborate
fitting functions. There are also other interesting attempts to extract 
useful information using finite size effects in ChPT \cite{Bie03}.

Given the difficulties associated with understanding chiral singularities
in a realistic calculation of QCD, in this paper we explore the subject
in strong coupling lattice QCD with staggered fermions. We use a very 
efficient algorithm discovered recently to solve this model in the chiral 
limit \cite{Ada03}. Although the strong coupling 
limit suffers from severe lattice artifacts, when the quarks are massless 
lattice QCD with staggered fermions has an exact $O(2)$ chiral symmetry which
is broken spontaneously. Thus, our model contains some of the
remnant physics of chiral singularities expected in QCD. In particular, 
there are light pions and it would be useful to understand the range of 
quark masses where the singularities predicted by conventional ChPT
can be seen.
 
Instead of focusing on chiral singularities that are logarithmic, in
this work we focus on power-like singularities that arise at finite
temperatures. Recently, it was shown with high precision that our
model undergoes a second order chiral phase transition at a critical
temperature $T_c$. This transition belongs to the $O(2)$ universality 
class \cite{Cha03}.  Thus, at a fixed temperature below $T_c$, within 
the the $O(2)$ scaling window, the long distance physics of our model
is described by a three-dimensional $O(2)$ field theory in its broken
phase. At this temperature there is a range of quark masses where
the light pions are describable by the conventional continuum ChPT. 
The effective action can be written in terms of $\vec{S}$, an $O(2)$ vector 
field with the constraint 
$\vec{S}\cdot\vec{S}=1$. At the lowest order this is given by 
\cite{Has90}
\begin{equation}
S_{\rm eff} = \int d^3 x 
\Big[\ \frac{F^2}{2} \ \partial_\mu \vec{S} \cdot \partial_\mu \vec{S} + 
 \Sigma \ \vec{h} \cdot \vec{S} \Big]
\label{chlag}
\end{equation}
where $\vec{h}$ is the magnetic field and $|\vec{h}|$ is identified with 
the quark mass. 
The two low energy constants $F^2$ and $\Sigma$ that appear at this order are 
the pion decay constant and the chiral condensate respectively. We note 
that since we are discussing a three-dimensional effective theory, $F^2$ 
has the dimensions of inverse length. Thus, we can define a correlation
length $\xi\equiv 1/F^2$. For massless quarks $\xi$ diverges as 
$(T_c-T)^{-\nu}$ at $T_c$. In order to connect our model with ChPT and
avoid lattice artifacts we need to choose the temperature and quark masses
such that $1 << \xi << 1/M_\pi$ (lengths are measured in lattice units) where
$M_\pi$ is the pion mass. In this region we expect observables such as 
the chiral condensate, the pion mass and the pion decay constant satisfy 
the expansion
\begin{equation}
\langle O \rangle =  z_0 + z_1\sqrt{m} + z_2 m + z_3 m \sqrt{m} + ...,
\end{equation}
where the $\sqrt{m}$ behavior is the power-like singularity arising due
to the infrared pion physics \cite{Wal75}. The main objective of this paper 
is to detect these singularities in the context of strong coupling lattice 
QCD with staggered fermions. Such power-like singularities have been observed 
in spin models \cite{Eng01}, but as far as we know have not been studied with 
precision in lattice QCD calculations. 

\section{The Model and Observables}

The partition function of the model we study in this article is given by
\begin{equation}
Z(T,m) = \int [dU] [d\psi d\bar\psi]\ \exp\left(-S[U,\psi,\bar\psi]\right),
\label{uNpf}
\end{equation}
where $[dU]$ is the Haar measure over $U(3)$ matrices and 
$[d\psi d\bar\psi]$ specify Grassmann integration. At strong coupling, the 
Euclidean space action $S[U,\psi,\bar\psi]$ is given by
\begin{equation}
\label{fact}
- \sum_{x,\mu} \frac{\eta_{x,\mu}}{2}\Big[\bar\psi_x U_{x,\mu} 
\psi_{x+\hat{\mu}}
- \bar\psi_{x+\hat{\mu}} U^\dagger_{x,\mu} \psi_x\Big]
- m \sum_x \bar\psi_x\psi_x,
\end{equation}
where $x$ refers to the lattice site on a periodic four-dimensional 
hyper-cubic lattice of size $L$ along the three spatial directions 
and size $L_t$ along the euclidean time direction. The index 
$\mu=1,2,3,4$ refers to the four space-time directions, $U_{x,\mu} \in U(3)$ 
is the usual links matrix representing the gauge fields, and 
$\psi_x,\bar\psi_x$ 
are the three-component staggered quark fields. The gauge fields satisfy 
periodic boundary conditions while the quark fields satisfy either periodic 
or anti-periodic boundary conditions. The 
factors $\eta_{x,\mu}$ are the well-known staggered fermion phase factors. 
We choose them to be
$\eta_{x,\mu}^2 = 1, \mu=1,2,3$ (spatial directions) and $\eta_{x,4}^2 = T$ 
(temporal direction), where the real parameter $T$ acts 
like a temperature. By working on asymmetric lattices with $L_t << L$ at
fixed $L_t$ and varying $T$ continuously one can study finite temperature 
phase transitions in strong coupling QCD \cite{Boy92}.
We use $U(3)$ gauge fields 
instead of $SU(3)$ in order to avoid inefficiencies in the algorithm due 
to the existence of baryonic loops. This distinction is not
important for our study since the baryons are expected to have a mass 
close to the cutoff.

The partition function given in Eq.~(\ref{uNpf}) can be rewritten in a 
monomer-dimer representation as discussed in detail in \cite{Ros84,Cha03}. 
Every configuration in the new representation
is described by monomer variables $n_x=0,1,2,3$ associated to the sites, 
and dimer variables  $b_{x,\mu}=0,1,2,3$ associated to the bonds
connecting neighboring sites $x$ and $x+\hat{\mu}$, along with the 
constraint that at each site, 
$n_x + \sum_\mu [b_{x,\mu} + b_{x-\hat{\mu},\mu}] = 3$.

In order to study chiral physics we focus on the following observables:
\begin{itemize}
\item[(i)] The chiral condensate
\begin{equation}
\langle\phi\rangle = \frac{1}{L^3} \frac{1}{Z}\frac{\partial}{\partial m} 
Z(T,m),
\end{equation}
\item[(ii)] the chiral susceptibility
\begin{equation}
\chi = \frac{1}{L^3} \frac{1}{Z}\frac{\partial^2}{\partial m^2} Z(T,m),
\end{equation}
\item[(iii)] the helicity modulus
\begin{equation}
Y = \frac{1}{L^3}\Bigg\langle 
\Big\{[\sum_x J_{x,1}]^2 + [\sum_x J_{x,2}]^2 + [\sum_x J_{x,3}]^2 \Big\}
\Bigg\rangle,
\end{equation}
where $J_{x,\mu} = \sigma_x (b_{x,\mu} - N/8)$, with $\sigma_x = 1$ on
even sites and $\sigma_x = -1$ on odd sites and 
\item[(iv)] the pion mass, obtained using the exponential decay of the
correlation function along one of the spatial directions $\mu=1,2,3$:
\begin{equation}
\lim_{|x_\mu| \rightarrow \infty} 
\sum_{x_\perp} \langle \sigma_x n_x n_0 \rangle = 
C \mathrm{e}^{-M_\pi |x_\mu|},
\end{equation}
\end{itemize}
where $x_\perp$ refers to components of the coordinate $x$ perpendicular to 
the $\mu$ direction.
When $m=0$ the current $J_{x,\mu}$ is the conserved current associated with 
the  $O(2)$ chiral symmetry. As discussed in \cite{Has90}, one can define
the pion decay constant at a quark mass $m$ to be 
$F_m^2 \equiv \lim_{L\rightarrow \infty} Y$. For $m=0$ we then obtain
$F = \lim_{m\rightarrow 0} F_m$, the pion decay constant introduced 
in Eq.(\ref{chlag}). We can also define 
$\Sigma_m = \lim_{L\rightarrow \infty} \langle \phi\rangle$ the infinite 
volume chiral condensate. Again $\Sigma = \lim_{m\rightarrow 0} \Sigma_m$.

\section{Results}

\begin{figure}[htb] 
\vskip0.3in
\begin{center}
\includegraphics[width=0.45\textwidth]{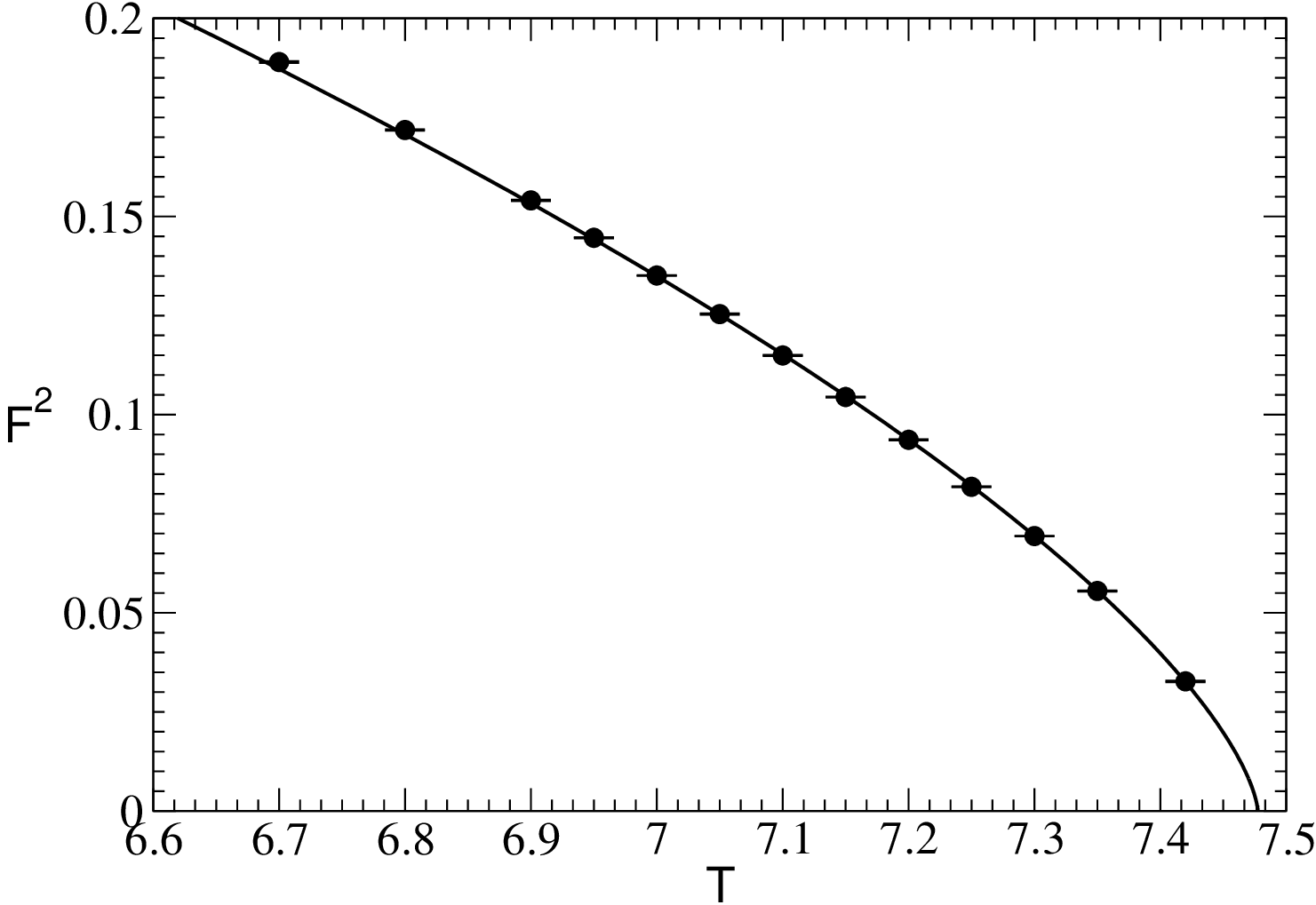}
\end{center}
\caption{\label{fig1}
Plot of $F^2$ vs. $T$. The solid line represents the function 
$0.2217(7.47739-T)^{0.67155}$.}
\end{figure}

We have done extensive computations at $L_t=4$. 
It has been shown with high precision in \cite{Cha03} that this model
undergoes a chiral phase transition which belongs to the three-dimensional 
$O(2)$ universality class at a critical temperature $T_c=7.47739(3)$. 
Figure \ref{fig1} shows the plot of $F^2$ as a function of $T$. 
Based on universality we expect $\xi^{-1} \equiv F^2 = C (T_c-T)^\nu$ for 
$T<T_c$ with $\nu = 0.67155(27)$ \cite{Cam01}. If we fit our results to 
this form in the range $7.05 \leq T \leq 7.42$, with $T_c$ and $\nu$ fixed 
to the expected values, we find $C=0.2217(1)$ with $\chi^2/$DOF$=1.03$. 
Including $T=7.0$ in the fit makes the $\chi^2$/DOF jump to $2.0$.
At a temperature very close to $T_c$, we expect 
$\xi$ to be extremely large and it would be difficult to satisfy 
$1/M_\pi >> \xi$ with our limited computing resources.
On the other hand in order to avoid lattice artifacts it is important
not to have $\xi \sim 1$. Using the above analysis we estimate that $T=7.0$ is 
at the edge of the $O(2)$ scaling window and hence we choose to fix $T$ at 
this value and study the long distance physics near the chiral limit. 
We vary the spatial lattice size from $L=8$ to $L=96$ for a variety of masses 
$0 \leq m \leq 0.025$. At $m=0$ we also have data at $L=120,144$. Below 
we discuss our main results.

\subsection{The Pion Decay Constant}

Our results show that for $L \geq 32$ the finite size effects on $Y$ are 
smaller than
the statistical errors. Hence it is relatively easy to 
compute $F_m^2$ by extrapolating results for $Y$ at different volumes. 
In Table \ref{tab1} we give the results for some values of the quark mass.
\begin{table}[htb]
\begin{tabular}{|c|c||c|c|}
\hline
$m$      & $F_m$     & $m$      & $F_m$ \\ \hline
0.00000  & 0.3675(1)   & 0.0050  & 0.4063(1)  \\
0.00025  & 0.3703(1)   & 0.0075  & 0.4188(1)  \\
0.00050  & 0.37300(15) & 0.0100  & 0.42890(15) \\
0.00100  & 0.37804(15) & 0.0200  & 0.4583(2) \\
0.0025   &  0.3903(2)  & 0.0250  & 0.4693(2) \\
\hline
\end{tabular}
\caption{\label{tab1} $F_m$ as a function of $m$.}
\end{table}
The first four terms in the chiral expansion of $F_m$ in three-dimensional 
$O(N)$ ChPT are given by \cite{Has90}
\begin{equation}
F_m = F[1 + a_1 \sqrt{m} + a_2 m + a_3 m\sqrt{m}],
\label{fpi}
\end{equation}
where $a_1 \propto (N-2)$. Since in our case $N=2$ we expect $a_1=0$.
\begin{figure}[htb]
\vskip0.3in
\begin{center}
\includegraphics[width=0.45\textwidth]{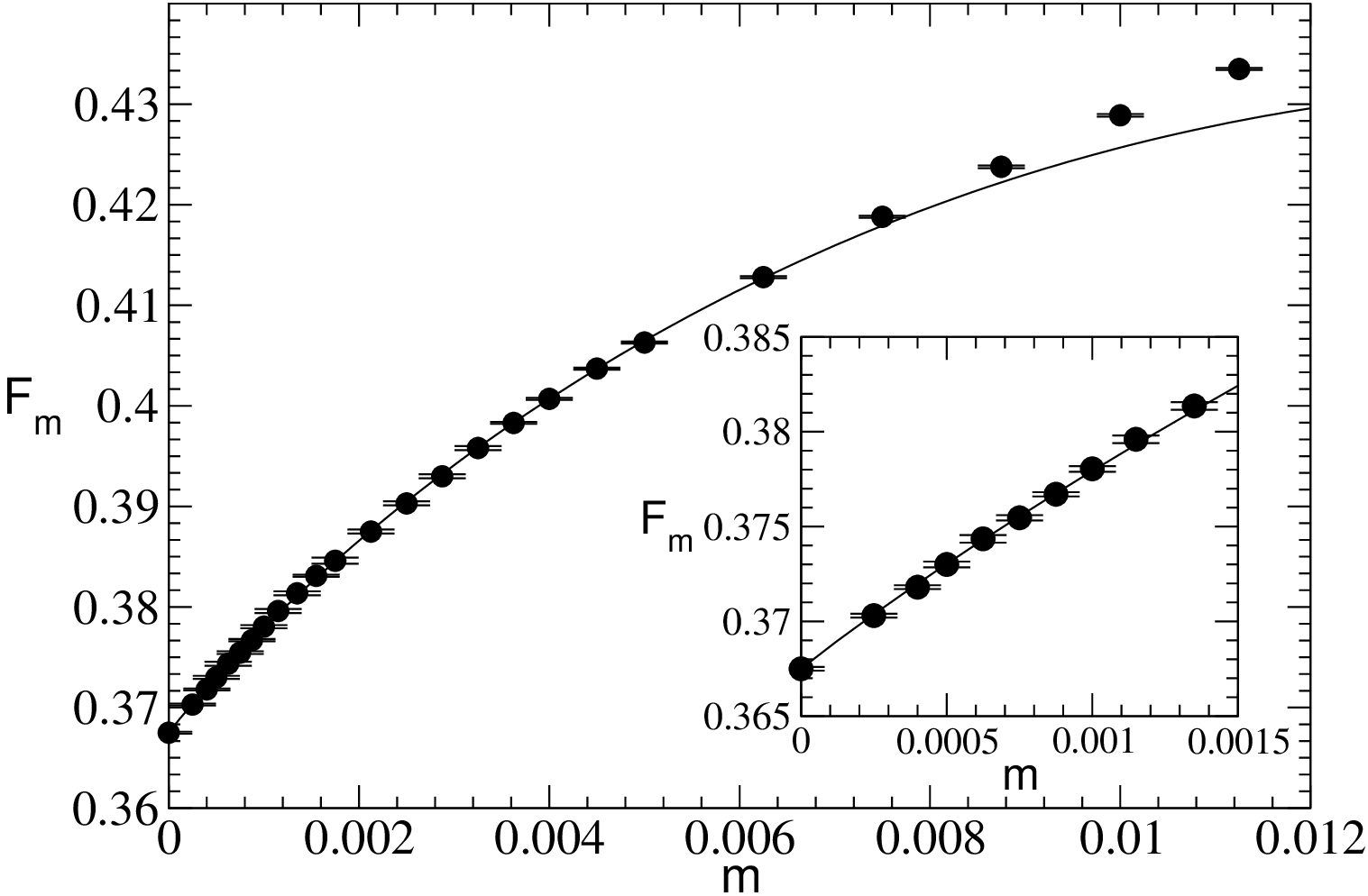}
\end{center}
\caption{\label{fig2}
Plot of $F_m$ vs. $m$. The solid line represents the function
$F_m=0.36747[1+34.2m - 185m\sqrt{m}]$.}
\end{figure}
In order to check if our results fit the predictions of ChPT 
we fit our data to Eq.(\ref{fpi}) in the
range $0 \leq m \leq 0.00625$. We find $F=0.3674(1)$, $a_1=0.01(2)$, 
$a_2=34.0(7)$ and $a_3 =-182(6)$ with $\chi^2$/DOF$=1.2$. 
Our data and the fit are shown in Fig. \ref{fig2}.
The prediction of ChPT that $a_1=0$ is in excellent 
agreement with our results. Fixing $a_1 = 0$ in the fit yields 
$F=0.36747(6)$, $a_2=34.3(3)$ and $a_3 =-185(3)$, while the
$\chi^2$/DOF remains essentially unchanged. Interestingly if
we fit the data in the range $0 \leq m \leq 0.00135$ with $a_1=a_3=0$ 
fixed, we find that $F=0.36764(7)$ and $a_2=28.3(3)$ with
$\chi^2$/DOF$=0.9$. This shows that there are systematic errors in 
evaluating the fitting parameters due to contamination from higher 
order terms. 
In the inset of Fig. \ref{fig2} we focus on the extremely small mass 
region and show 
that $a_1=0$ is even clear to the eye. This is one of the 
main results of our paper. We suggest that $a_1=0$ is a useful 
signature of the $O(2)$ universality and could be used in future 
studies of lattice QCD with staggered fermions.

\subsection{The Chiral Condensate}

Chiral perturbation theory predicts that the first four terms in the 
chiral expansion of $\Sigma_m$ are given by 
\begin{equation}
\Sigma_m = \Sigma [1 + b_1 \sqrt{m} + b_2 m + b_3 m\sqrt{m}].
\label{cc}
\end{equation}
As shown in \cite{Has90}, $b_1 \propto (N-1)$, implying that $b_1 \neq 0$ 
in our case. In order to test how well Eq. (\ref{cc}) describes our 
model, we compute $\Sigma_m$ at small masses.

At $m=0$ we compute $\Sigma$ by using the finite size scaling formula for $\chi$ 
given by \cite{Has90}
\begin{equation}
\chi = \frac{1}{N}\Sigma^2 L^3\Big[ 1 + 0.226 (N-1) \frac{1}{F^2 L} + 
\frac{\alpha}{L^2} +...\Big],
\label{chptchi}
\end{equation}
where $\alpha$ depends on the higher order low energy constants of the 
chiral Lagrangian. When $F$ is fixed to $0.3675$ obtained earlier, our
results for $16 \leq L \leq 144$ fit very well to this formula. We find
$\Sigma = 2.2648(10)$, $\alpha=4.6(3)$ with $\chi^2$/DOF$=0.87$.

\begin{table}[htb]
\begin{tabular}{|c|c||c|c|}
\hline
$m$      & $\Sigma_m$   & $m$     & $\Sigma_m$ \\ \hline
0.00000  & 2.2648(10)   & 0.0050  & 2.6340(08) \\
0.00025  & 2.2340(10)   & 0.0075  & 2.7278(15) \\
0.00050  & 2.3560(10)   & 0.0100  & 2.8025(15) \\
0.00100  & 2.4040(05)   & 0.0200  & 3.0199(11) \\
0.0025   & 2.5103(08)   & 0.0250  & 3.0986(08) \\
\hline
\end{tabular}
\caption{\label{tab2} Values of $\Sigma_m$ at selected values of $m$.}
\end{table}

\begin{figure}[htb]
\vskip0.3in
\begin{center}
\includegraphics[width=0.45\textwidth]{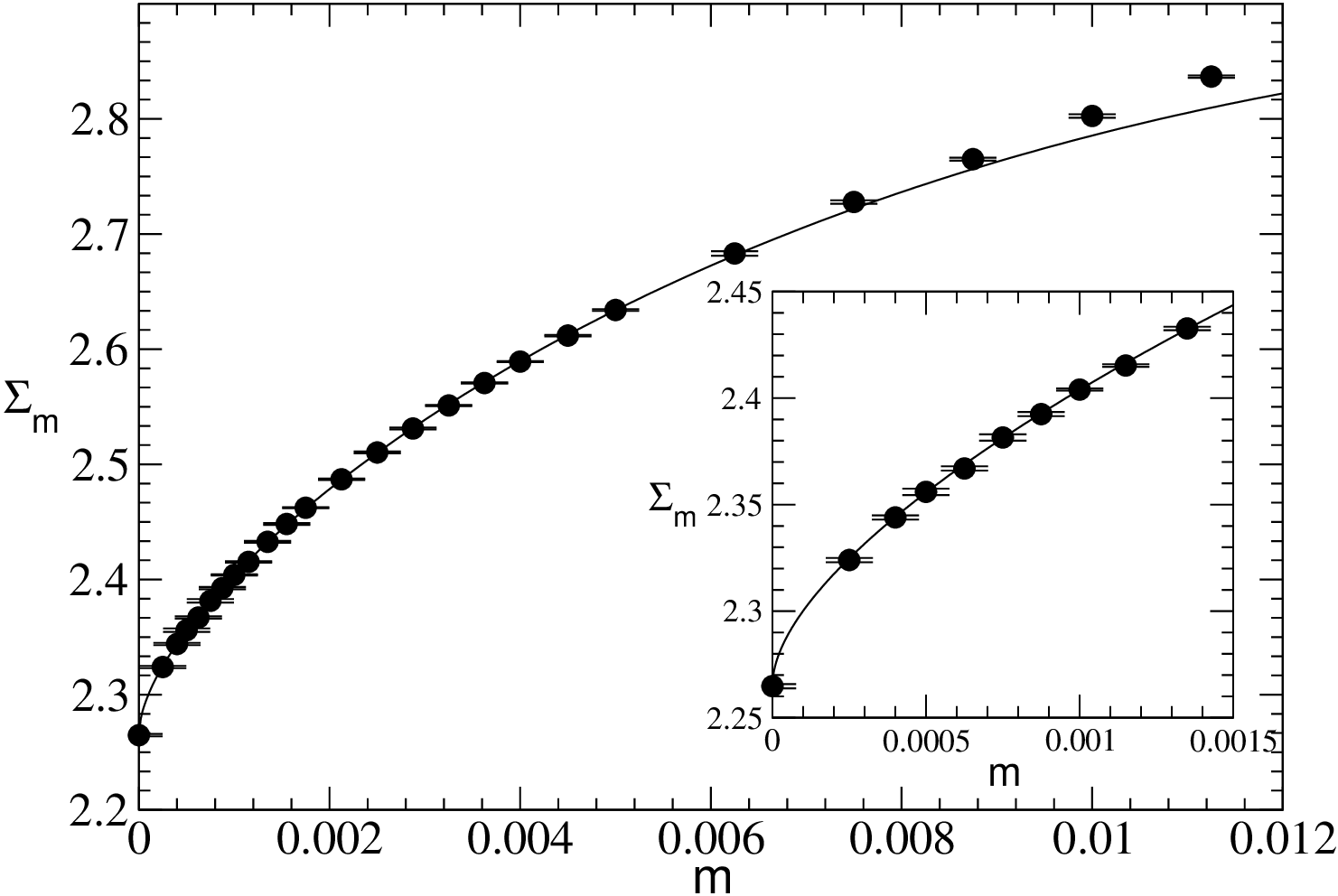}
\end{center}
\caption{\label{fig3} Plot of $\Sigma_m$ vs. $m$. The solid line
represents the function $\Sigma_m = 2.2642[1+1.36\sqrt{m} + 
23 m - 135 m\sqrt{m}]$.}
\end{figure}
  
For $m\neq 0$ one can measure $\Sigma_m$ by extrapolating the results
for $\langle \phi \rangle$ at different volumes. However, due to
critical slowing down it is difficult to measure the condensate accurately 
at a fixed small mass on large volumes. With our computing resources we find 
that this procedure gives reliable answers for $\Sigma_m$ only up to 
$m\sim 0.001$. On the other hand when $m\neq 0$, in the large volume limit
we expect $\chi \sim \Sigma_m^2 L^3$ up to an additive constant plus 
exponentially  small corrections. We find that the signal for $\Sigma_m$ 
extracted from fitting
the data to this form is much cleaner at small masses than the signal
obtained from the direct measurement. Further, the values of $\Sigma_m$ 
obtained using this procedure agree well with the direct measurement at 
larger quark masses. In table \ref{tab2} we give the values of $\Sigma_m$ 
obtained for selected values of the quark masses.
Fitting our results to Eq.(\ref{cc}) in the region $0 \leq m \leq 0.00625$ 
gives $\Sigma=2.2642(10)$, 
$b_1=1.36(4)$, $b_2 = 23(1)$, $b_3 = -135(10)$ with a $\chi^2$/DOF$=1.1$.
Our results are shown in Fig \ref{fig3}.
We note that we cannot find a mass range within our results
in which we can find a good fit when we fix $b_2=b_3=0$. However, 
in the range $0 \leq m \leq 0.00175$ we can set $b_3=0$
to obtain $\Sigma=2.2643(10)$, $b_1=1.44(3)$ and 
$b_2=15.2(7)$ with a $\chi^2$/DOF$=1$. Note that $b_2$
changes by about 30\% when this different fitting procedure is
used, while $b_1$ is more stable.

\begin{table}[htb]
\begin{tabular}{|c|c||c|c|}
\hline
$m$      & $\Sigma_m$  & $m$     & $\Sigma_m$ \\ \hline
0.00000  & 0.000       & 0.0050  & 0.2812(8) \\
0.00025  & 0.0650(3)   & 0.0075  & 0.3399(5) \\
0.00050  & 0.0920(4)   & 0.0100  & 0.3871(4) \\
0.00100  & 0.1295(4)   & 0.0200  & 0.5285(10) \\
0.0025   & 0.2022(3)   & 0.0250  & 0.5837(8) \\
\hline
\end{tabular}
\caption{\label{tab3} Values of $M_\pi$ at selected values of $m$.}
\end{table}

\begin{figure}[htb]
\vskip0.3in
\begin{center}
\includegraphics[width=0.45\textwidth]{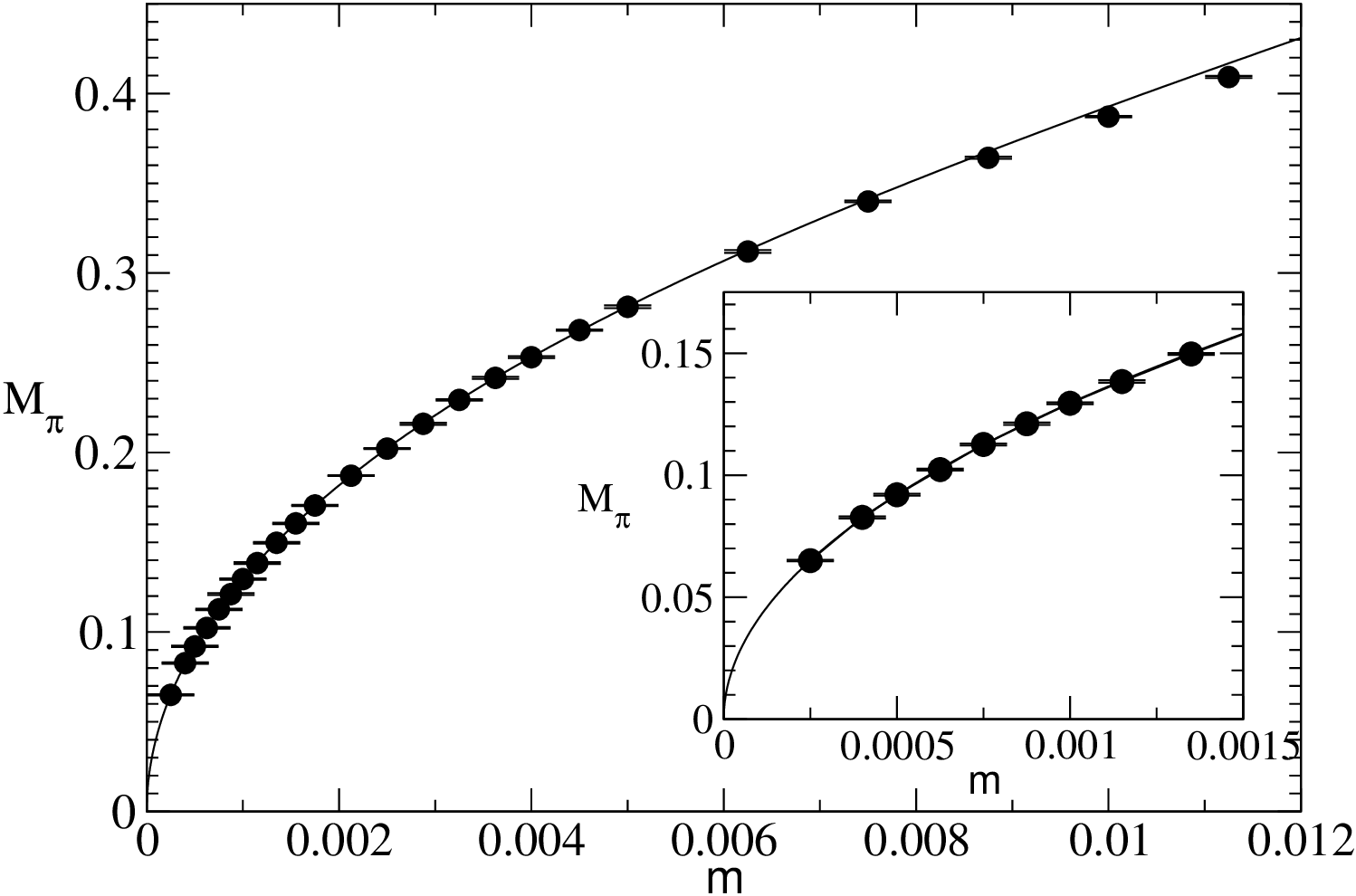}
\end{center}
\caption{\label{fig4} Plot of $M_\pi$ vs. $m$. The solid line
represents the function 
$M_\pi = 4.095 \sqrt{m}[1+0.68 \sqrt{m} - 25.8 m + 149 m \sqrt{m}]$.}
\end{figure}

\subsection{The Pion Mass}

The first four terms in the chiral behavior of the pion mass are 
predicted to be of the form 
\begin{equation}
M_\pi = \sqrt{\frac{\Sigma}{F^2}}\sqrt{m}
\Big[1 + c_1 \sqrt{m} + c_2 m + c_3 m\sqrt{m}\Big].
\label{mpi}
\end{equation}
Further, the chiral Ward identities imply that $c_1 = b_1/2 - a_1$.
Our estimates for the pion masses for a selected range of $m$ are shown
in Table \ref{tab3}.
Fitting our data to Eq.(\ref{mpi}) after fixing $\Sigma=2.2648$ 
and $F=0.3675$ obtained above, we find $c_1 = 0.55(16)$,
$c_2=-21(6)$ and $c_3 =110(50)$ with $\chi^2/$DOF$=0.5$. We see that 
our data is consistent with the relation $c_1=b_1/2 - a_1$ 
although the error in $c_1$ is large. Fixing $c_1=b_1/2=0.68$ obtained
from fitting the chiral condensate,
yields $c_2=-25.8(8)$ and $c_3=149(14)$ without changing the
quality of the fit. This latter fit along with our results for $M_\pi$
are shown in Fig. \ref{fig4}.

\section{DISCUSSION}

It is expected that the natural expansion parameter for ChPT 
in three dimensions is $x\equiv M_\pi/(4\pi F^2)$. Let us find
the values of $x$ where the chiral expansion up to 
a certain power of $x$ is sufficient to describe the data reasonably. 
Of course this question is model-dependent, since some models may have larger 
contributions at higher orders compared to others. Here we ask this 
question in the context of the model studied in this paper.

\begin{figure}[htb]
\vskip0.3in
\begin{center}
\includegraphics[width=0.45\textwidth]{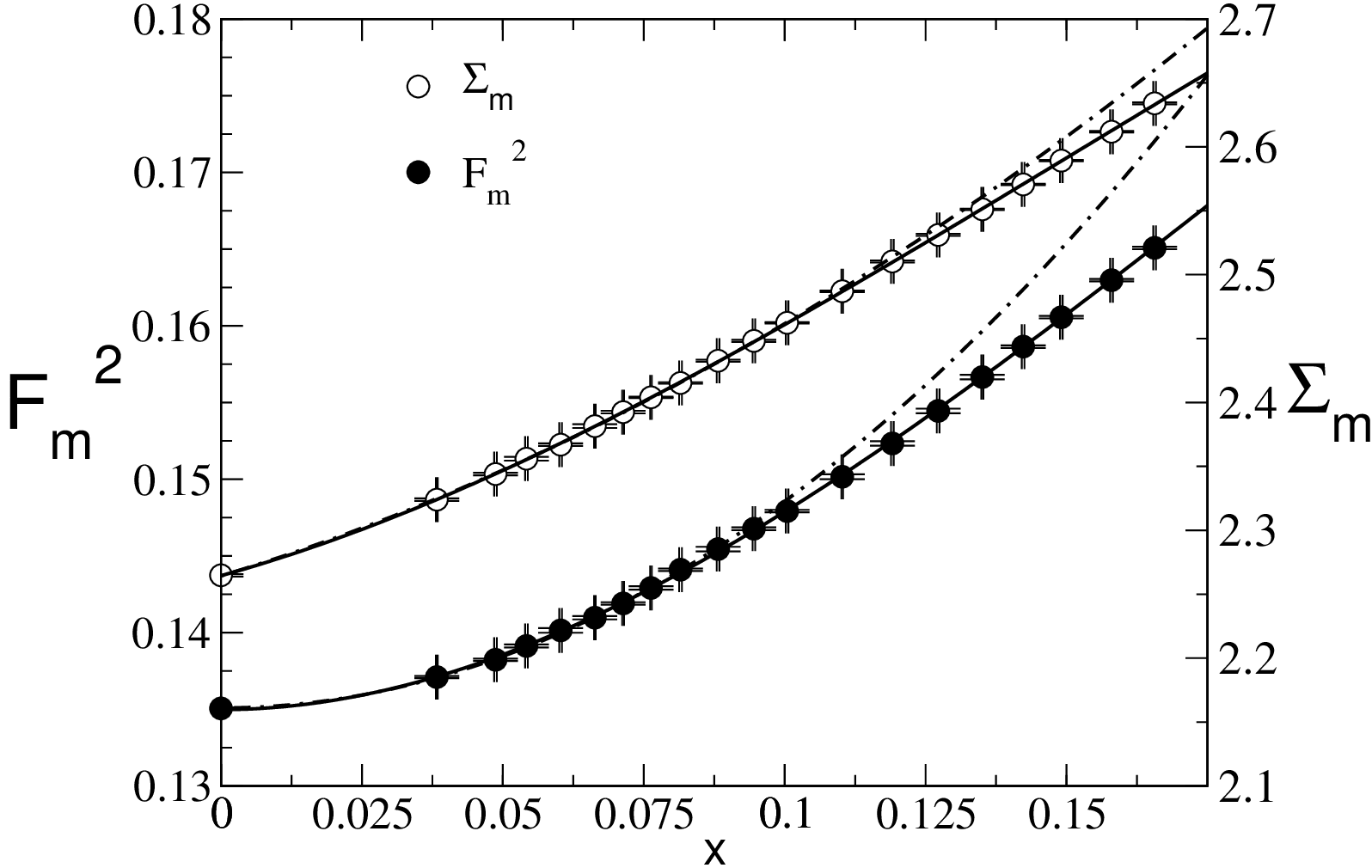}
\end{center}
\caption{\label{fig5} Plot of $F^2_m$ and $\Sigma_m$ vs. 
$x\equiv M_\pi/(4\pi F^2)$. The solid lines and dot-dashed lines are 
fits to two different orders of chiral expansion discussed in the text.}
\end{figure}

In order to understand the dependence of $F^2_m$ and $\Sigma_m$ on
$x$ we plot in Fig. \ref{fig5} these two observables as functions of $x$. 
The solid lines are the best 
fits to the form $z_0 + z_1 x + z_2 x^2 + z_3 x^3$, while the dot-dashed 
lines are the best fits to a smaller range in $x$ with $z_3$ set to $0$. 
As the graph indicates the error in $\Sigma_m$ due to the absence of
the $z_3$ term is different from the error in $F_m^2$ at a given value of $x$. 
In order to determine $F_m^2$ within say $5\%$ we need the $z_3$ 
term even at $x \sim 0.15$. This shows that our model contains 
important higher order terms. An interesting 
question which we cannot answer at this point is whether this property is 
generic or not. In any case we have shown that the connection 
of lattice QCD data to ChPT is indeed possible as long 
as one does not assume that higher order terms are negligible. 
For the lowest order terms to dominate it may be necessary to
go to much smaller quark mases than has been possible until now. 
In fact it is much easier to connect our results with ChPT at $m=0$, which is 
called the $\epsilon$-regime of ChPT \cite{Bie03}. 
Thus, finding an algorithm to work directly at $m=0$ is a useful goal 
to strive for in the future. Finally we hope that our results will motivate 
further work to uncover the power-law chiral singularities at finite 
temperatures in Lattice QCD at weaker couplings. This should be easier
than looking for logarithmic singularities at zero temperature.

\section*{Acknowledgments}

We thank M. Golterman, S. Hands, T. Mehen, S. Sharpe, R. Springer and 
U.-J. Wiese for helpful comments. This work was supported in part by the 
Department of Energy (DOE) grant DE-FG-96ER40945. SC is also supported 
by an OJI grant DE-FG02-03ER41241. The computations were performed on the 
CHAMP, a computer cluster funded in part by the DOE and located in 
the Physics Department at Duke University.

\end{document}